# CHEMICAL AND FORENSIC ANALYSIS OF JFK ASSASSINATION BULLET LOTS: IS A SECOND SHOOTER POSSIBLE?

BY CLIFF SPIEGELMAN, WILLIAM A. TOBIN, WILLIAM D. JAMES,
SIMON J. SHEATHER, STUART WEXLER AND D. MAX ROUNDHILL

*Texas A&M University, Forensic Engineering International,
Texas A&M University, Texas A&M University,
Hightstown High School and Chem Consulting*

The assassination of President John Fitzgerald Kennedy (JFK) traumatized the nation. In this paper we show that evidence used to rule out a second assassin is fundamentally flawed. This paper discusses new compositional analyses of bullets reportedly to have been derived from the same batch as those used in the assassination. The new analyses show that the bullet fragments involved in the assassination are not nearly as rare as previously reported. In particular, the new test results are compared to key bullet composition testimony presented before the House Select Committee on Assassinations (HSCA). Matches of bullets within the same box of bullets are shown to be much more likely than indicated in the House Select Committee on Assassinations' testimony. Additionally, we show that one of the ten test bullets is considered a *match* to one or more assassination fragments. This finding means that the bullet fragments from the assassination that match could have come from three or more separate bullets. Finally, this paper presents a case for reanalyzing the assassination bullet fragments and conducting the necessary supporting scientific studies. These analyses will shed light on whether the five bullet fragments constitute three or more separate bullets. If the assassination fragments are derived from three or more separate bullets, then a second assassin is likely, as the additional bullet would not easily be attributable to the main suspect, Mr. Oswald, under widely accepted shooting scenarios [see Posner (1993), *Case Closed*, Bantam, New York].

**1. Introduction.** The assassination of President John Fitzgerald Kennedy (JFK) was arguably the most traumatic event to the nation in the early 1960s. The assassination precipitated much scientific investigation and consequent testimony before Congressional subcommittees. The scientific investigations included a National Academy of Science Committee on Ballistic Acoustics report, forensic evidence on bullet fragment matching by the FBI crime lab and by Dr. Vincent P. Guinn, as well as more typical forensic evidence such as fingerprints, ballistic evaluations and simulations, crime scene investigations and reconstructions, and autopsy. The House Select Committee on Assassinations [U.S. Cong. House (1979)] concluded that there was a probable conspiracy, but one in which an additional shooter (likely firing from the Grassy Knoll) missed all limousine occupants. The evidentiary anchor for the second part of that conclusion—that any additional shot must have missed—was Dr. Vincent P. Guinn's comparative bullet lead analysis [U.S. Cong. House (1979)]. It demonstrated to the Committee, with the added insights from ballistics testing, that all of the recovered ballistics material had only one common origin, Mannlicher–Carcano rounds fired from Oswald's rifle.

The $+/-$ numbers in Table 1 represent standard deviations based upon a single observation using Poisson counting statistics. The exception is the last row of the table where 3 fragments constitute sample CE 840, and Dr. Guinn (1979) analyzed all of them. In that case, the standard deviation measurement was constructed from 3 replicate measurements. Dr. Guinn grouped the bullet fragments listed in Table 1 [Guinn (1979)] into two groups. Fragments CE 399 and CE 842 formed one group and fragments CE 567, CE 843 and CE 840 formed the second group. Guinn testified [U.S. Cong. House (1979)] that only two bullets represented the sources of those fragments.

We compare data from our studies of other JFK assassination-related batch bullets to selected contextual excerpts of Dr. Guinn's House Select Committee on Assassinations' testimony [U.S. Cong. House (1979)] to the recent National Research Council (NRC) report *Forensic Analysis*: *Weighing Bullet Lead Evidence* (2004) and, also, to author experience in compositional analysis of bullet lead. We will show that Dr. Guinn's most important testimony to the House Select Committee on Assassinations [U.S. Cong. House (1979)] is unlikely justified. In particular, we shall show that random matches to assassination fragments of bullets from the same box are not as rare as Dr. Guinn testified. Most importantly, our studies and analyses of individual bullet compositions, bullet lead source compositions and compositional mixtures in packaged retail boxes show that Dr. Guinn's statements about the uniqueness of individual bullets from the brand of bullets believed to be used in the assassination are seriously flawed.

TABLE 1
*Guinn's NAA results for silver and antimony in bullets and fragments*

| Specimen | Description | Silver, ppm | Antimony, ppm |
|---|---|---|---|
| CE 339 (Q1) | Bullet from stretcher | $8.8 \pm 0.5$ | $833 \pm 9$ |
| CE 842 (Q9) | Largest metal fragment from Connally's arm | $9.8 \pm 0.5$ | $797 \pm 7$ |
| CE 567 (Q2) | Bullet fragment from front seat cushion | $8.1 \pm 0.6$ | $602 \pm 4$ |
| CE 843 (Q4) | Larger metal fragment from the President's head | $7.9 \pm 0.3$ | $621 \pm 4$ |
| CE 840 (Q14) | Metal fragments from rear floorboard carpet | $8.2 \pm 0.4$ | $642 \pm 6$ |

## 2. Forensic issues.

2.1. *Deciding whether bullets "match."* Before describing the experimental technique used for our study, we first explain how bullet compositions are associated as forensic "matches." A match criterion establishes a rule for declaring bullet compositions close enough to be considered analytically indistinguishable. Typically, a "match" criterion is established, such as confidence interval overlap between mean analyte levels (chemical element levels in the lead matrix) measured in two different specimens. The confidence intervals are typically based upon multiples of estimated standard deviations from replicate measurements (called a multiple of sigmas method). Since 1993, the analytes of two or more bullet specimens were compared as described below. A typical analysis, which we consider here, compares the "questioned" or crime scene evidence of "unknown" origin with samples "known" to have been in the suspect's possession. The chemical elements present in bullet lead that were most frequently analyzed after approximately 1989 and used for forensic comparisons were antimony, copper, arsenic, bismuth, silver, tin and cadmium. In earlier years (roughly from the 1960s to 1993), only antimony, copper, arsenic and sometimes silver were characterized. The fewer elements that are used to declare a match the more likely that matches, including coincidental ones, will occur [see NRC report (2004)]. When an analyte concentration in bullets from a crime scene specimen and in those from a suspect bullet were compared and the concentrations fell within the established "match" criterion range, that particular elemental comparison was declared a "match." If all the remaining elemental comparisons between the crime scene specimen and "known" specimen also fell within the established "match" criterion, the two bullet specimens were then declared to be analytically "indistinguishable."

2.2. *The brand of bullets believed to be used in the Kennedy assassination.* In the House Select Committee on Assassinations hearing [U.S. Cong. House (1979)], Dr. Guinn testified about the brand of bullets believed to be used, namely, Winchester Cartridge Company Mannlicher–Carcano bullets, and about the five bullet fragments of forensic interest in the assassination. In particular, he concluded that "This great variation from bullet to bullet from the same box thus indicated that, for this unusual kind of ammunition, it would be possible to distinguish one bullet (or bullet fragment) from another, even though they both came from the same box of Mannlicher–Carcano cartridges" [U.S. Cong. House (1979)].

The authors purchased 2.5 boxes of Winchester Cartridge Company Mannlicher–Carcano bullets from 2 of the only 4 separate lots ever produced. One box came from lot 6000, and one and a half from lot 6003.[1] We then analyzed 10 bullets from each box. The measurement approach was similar to that used by Dr. Guinn except that we used more appropriate standards, a known quality control procedure, and analyzed physical samples having a known geometry. *One of the bullets analyzed matched an assassination fragment.* We also found that many bullets in the same box have matching antimony and silver levels; this discovery is contrary to Dr. Guinn's testimony that based on these two elements virtually every bullet of this type is unique. Peele et al. (1991) found similar supporting evidence of many bullets matching within boxes of bullets, including those of other large US bullet manufacturers. Further, we have found that about 14 percent of bullets in the "1837" bullet FBI empirical compilation reported in Spiegelman and Kafadar (2006) from murder cases have antimony concentrations at or below those of the assassination fragments. Approximately the same percentage of bullets is at or below the antimony concentrations of the Western Cartridge Company bullets measured by Dr. Guinn. The percentages vary somewhat with different FBI empirical compilations. For the other (non-"1837") FBI bullet compilations, it is not clear which case entries (samples) correspond to bullets rather than buckshot, other irrelevant entries and/or null cell values. Regardless of assumptions used to interpret various FBI bullet datasets, low antimony bullets are not rare. Our findings also call into question other aspects of Dr. Guinn's testimony [U.S. Cong. House (1979)], such as the following exchange that followed Dr. Guinn's testimony about the low antimony content of Western Cartridge Company Mannlicher–Carcano bullets:

> **Mr. Wolf:** In your professional opinion, Dr. Guinn, is the fragment removed from General Walker's house a fragment from a Western Cartridge Company Mannlicher–Carcano bullet?
> **Dr. Guinn:** I would say that it is extremely likely that it is, because there are very few, very few other ammunitions that would be in this range. I don't know of any that are specifically this close as these numbers indicate, but somewhere near them there are a few others, but essentially this is in the range that is rather characteristic of Western Cartridge Company Mannlicher–Carcano bullet lead.

---

[1]Preliminary background investigation revealed some uncertainty as to the source of 10 of the bullets that were purchased as a half box. After additional investigation and information from the bullet supplier (including photos), it is concluded likely that the 10 originated from lot 6003, and we denote this lot as 6003P.

There has never been a comprehensive or forensically meaningful statistically based sampling study of bullet composition by manufacturer, location or epoch.

Assuredly, the passage of time has allowed a more sophisticated analysis of the data obtained by Dr. Guinn. Our final point will be that it may be possible to provide substantial evidence of a third assassination bullet by following the guidelines in the NRC report (2004) and reanalyzing the assassination fragments. In any case, we believe that there is no scientific basis from the fragment matching performed by Dr. Guinn to conclude that only two bullets were the sources of the assassination fragments [U.S. Cong. House (1979)]:

> **Mr. Wolf:** What is the number of bullets, in your opinion?
> **Dr. Guinn:** These numbers correspond to two bullets. Two of the samples have indistinguishable compositions, indicating that they came from the same bullet, and the other three particles are evidently samples from another bullet.
> **Mr. Wolf:** So it is your opinion that the evidence specimens represent only evidence of two bullets, is that correct?
> **Dr. Guinn:** Yes, sir, there is no evidence for three bullets, four bullets, or anything more than two, but there is clear evidence that there are two.

Dr. Guinn may have been correct or incorrect about the number of bullets originating from the JFK fragments; the state of knowledge even today, but definitely about 30 years ago, remains too uncertain.

Finally, we note that much has been written about the JFK assassination, but most of the results such as those in Rahn and Sturdivan (2004) and Sturdivan and Rahn (2004), are based upon historical data using what we feel are inadequate models for bullet distribution and sample sizes that are too small. It is our belief that the lognormal model used by these authors to model antimony concentrations is one of a large set of reasonable models that might have been used. For example, the "distfit" command from the PLS_Toolbox for MATLAB ranks best-fitting distributions to the data using p-values from a chi-squared goodness-of-fit test. For the 14 data values that were used by Sturdivan and Rahn (2004), the exponential, Weibull, gamma, triangular and chi-squared distributions had p-values that were $0.71, 0.71, 0.70, 0.58$ and $0.54$, respectively, while the lognormal distribution had a p-value of $0.54$. Many other distributions, as well, are realistic possibilities for modeling the data, including Gumbel ($p = 0.45$) distributions. Based on the lognormal distribution, Sturdivan and Rahn (2004) concluded that there was a 2%–3% chance of a coincidental match. This result is clearly dependent on distributional assumptions and, hence, we believe that the 2%–3% figure is questionable. Sturdivan and Rahn (2004) are focused on chance matches within the whole population of Western Cartridge Company Mannlicher–Carcano bullets. We think a more relevant population for consideration is that consisting of bullets within the same box of Western

Cartridge Company Mannlicher–Carcano bullets. Comparing the composition of different bullet fragments is based on the assumption that different fragments from the same bullet are compositionally similar at all levels of sample size. With this issue in mind, Rahn and Sturdivan (2004) discuss fragment heterogeneity, based upon historical data from both Dr. Guinn and the FBI. Based on the FBI data, Rahn and Sturdivan (2004) found little heterogeneity between bullet fragments.

In our compositional analysis of 30 bullets, we found marginally detectable heterogeneity, which can be taken into account when matching bullets. We believe we have struck a good balance between the considerations adversely affecting compositional accuracy and precision relating to specimen size. Specimens that are too large result in self-absorption problems during analysis by NAA; specimens that are too small can result in analyses that may not be representative of the overall (bulk) composition, as demonstrated by Randich and Grant (2006).

The description of our measurement experiment can be found in the Appendix.

**3. Statistical analysis of Dr. Guinn's data.**   Dr. Guinn grouped the bullet fragments shown in Table 1 into two groups. Fragments CE 399 and CE 842 formed one group and fragments CE 567, CE 843 and CE 840 formed the second group. Guinn testified [U.S. Cong. House (1979)] that only two bullets represented the sources of those fragments. It is clear that CE 567 and CE 840 antimony measurements are $\pm 4$ estimated standard errors (of the mean) apart. Thus, we consider here a $\pm 4$ standard error method in order to remain consistent with Dr. Guinn's testimony. [A recent NRC report (2004) found that the standard procedure used by the FBI to match bullet fragments or bullets was based on $\pm 2$ standard deviations. When there are 4 replicate measurements the $\pm 4$ standard error matching method is the same as that used by the FBI.] Later in this paper we also consider a $\pm 2$ standard error matching criteria for matching our bullet measurements to Dr. Guinn's bullet fragment measurements. As Rahn and Sturdivan (2004) indicate, with the exception of CE 840, Dr. Guinn's measurements involved only one measurement *of each specimen*, so the estimated standard deviations and estimated standard errors are the same and are based upon Poisson counting statistics. The NRC report (2004) also discusses other matching techniques such as the equivalence t-test and a multivariate Hotelling's $T^2$ test.

Bias associated with any measurement system changes with time. To minimize this bias, it was common for the FBI crime lab to measure bullets or bullet fragments obtained from the crime scene and from a suspect within a short time frame. Dr. Guinn's measurements were made over a short time period and, hence, bias is not an issue for grouping his measurements. In

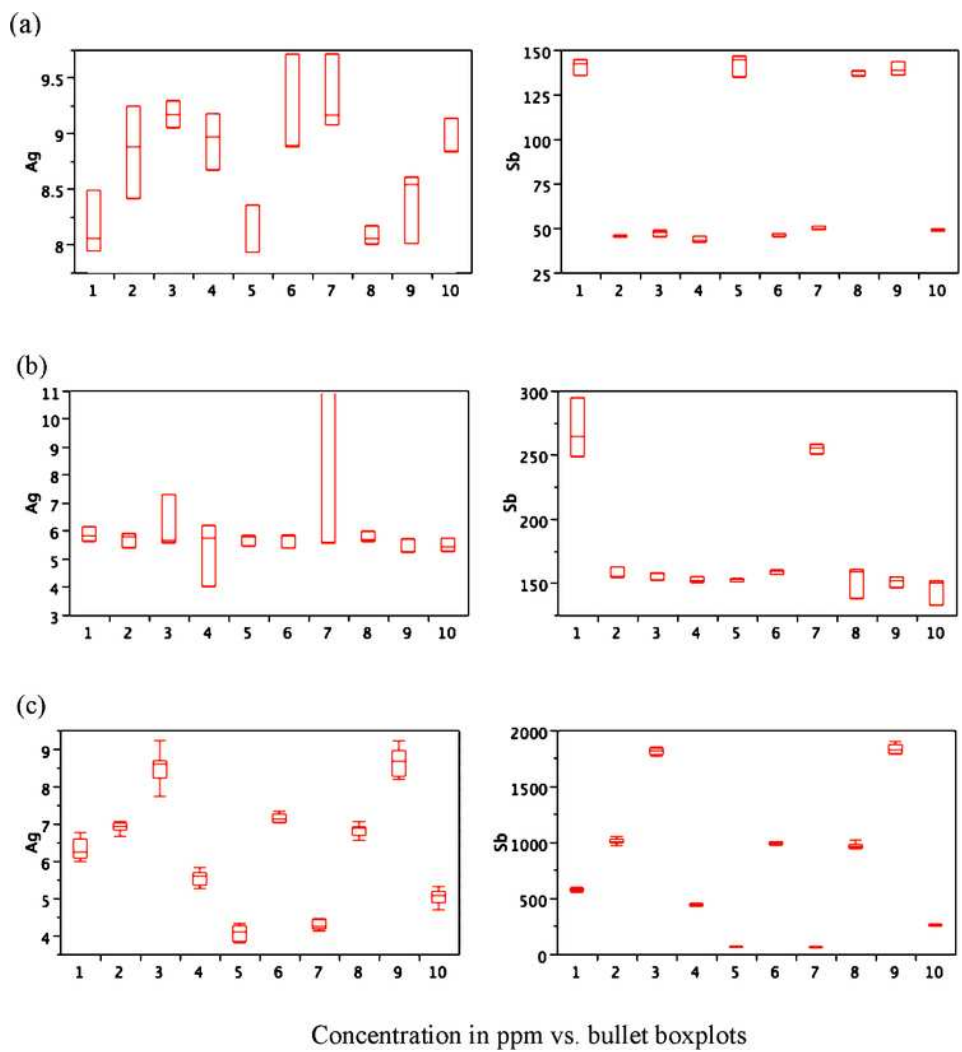

Concentration in ppm vs. bullet boxplots this report we included a bias correction for time as an additional component to the bullet comparisons. Based upon measurements on our three "standards," we have a mean offset from the certified values between 2% and 5.4%. Therefore, a bias correction to the antimony measurements that adds 2% to 5.4% is appropriate. This results in a match with one of our bullets with CE 567 that we will describe below. For the same reasons for silver, we added 5.5% to our measurements. As a result, the same bullet matches CE 567 at ±2 standard errors. Alternative approaches based on components of error adjustments are problematic, since Guinn's standard error estimates do not have traditional degrees of freedom associated with them. Dr. Guinn's uncertainty estimates do not take into account many sources of error, such as repositioning of samples, biases or instrument abnormalities. We invite readers who wish to use a different method for bias correction to consult our measurements on standards available in the online supporting material. Given the overlap in the boxplots in Figure 1 below, it is clear that many fragments corresponding to different Western Cartridge Company Mannlicher–Carcano bullets match on the two elements antimony and silver.

Fig. 1. (a) *Boxplots for silver and antimony measurements for lot 6000;* (b) *boxplots for silver and antimony measurements for lot 6003P;* (c) *boxplots of silver and antimony measurements for lot 6003.*

The NRC report (2004) recommends that seven elements be used for matching, and it is clear that many fragments that match on two elements may not match on three or more elements.

Of particular note is our bullet number one from lot 6003. We compare the outer, middle and inner fragments as well as the combined measurements to Guinn's measured fragment CE 567 (see Table 2), where means and standard errors are given.

Applying the criterion of ±4 standard errors, both the identified fragments and combined measurements from bullet 1 in lot 6003 would belong to Group 2 of Guinn's classification of fragments. The ±2-standard error match criterion fails, but that ignores the fact that there likely are biases in both Guinn's measurements and ours. As previously discussed, our biases are assessed to be approximately −5% to a positive 9.8% relative bias for silver and from −5.4% to −2% relative bias for antimony (data available within the online supporting documents). Once the adjustment for bias is performed, we can see that all fragments, labeled or not, have a ± two standard error match for antimony, but only the fragments labeled by position have a two standard error match for silver.

If every bullet ever manufactured, or if every bullet in a local geographic area, was of the same composition, there would be little or no forensic value of a claimed "match" or "nonmatch." As discussed in Tobin and Thompson (2006),

> Assuming you are satisfied that ... specimens were properly characterized as indistinguishable (therefore a "match" or "inclusion"), the next step for assessment of forensic significance involves estimation of probabilities for determination of probative value. ...there are two crucial questions: (1) how likely are the observed results if the samples had a common source; and (2) how likely are the observed results if the samples did *not* have a common source? Without valid answers to both questions, there is no way to assess the probative value of the forensic evidence for proving that the "matching" items had a common source.

TABLE 2
*Bullet # 1 lot 6003 measurements by radial location and combined giving summary statistics for antimony and silver*

| Location | Degrees of freedom (silver) | Silver, ppm (Mean ± standard error) | Silver bias range | Degrees of freedom (antimony) | Antimony, ppm (Mean ± standard error) | Antimony bias range |
|---|---|---|---|---|---|---|
| Outer fragment | 3 | $6.30 \pm 0.13$ | $-5.5\%$ to $+9.8\%$ | 3 | $578 \pm 9.75$ | $-5.4\%$ to $-2\%$ |
| Middle fragment | 2 | $6.66 \pm 0.05$ | $-5.5\%$ to $+9.8\%$ | 2 | $585 \pm 6.97$ | $-5.4\%$ to $-2\%$ |
| Inner fragment | 3 | $6.35 \pm 0.14$ | $-5.5\%$ to $+9.8\%$ | 3 | $581 \pm 7.56$ | $-5.4\%$ to $-2\%$ |
| All measurements for this bullet including unlabeled measurements | 19 | $6.30 \pm 0.06$ | $-5.5\%$ to $+9.8\%$ | 17 | $576 \pm 3.47$ | $-5.4\%$ to $-2\%$ |

Table 3
*Summary statistics for bullets 1, 8, 9 and 10 from lot 6003 by radial location for antimony and silver*

|    | Outer           | N | Middle          | N | Inner           | N | Whole bullet    | N  |
|----|-----------------|---|-----------------|---|-----------------|---|-----------------|----|
|    | Antimony (Sb)   |   |                 |   |                 |   |                 |    |
| 1  | $578 \pm 19.5$  | 4 | $585 \pm 12.1$  | 3 | $581 \pm 15.1$  | 4 | $576 \pm 3.47$  | 18 |
| 8  | $957 \pm 4.86$  | 3 | $952 \pm 17.4$  | 3 | $963 \pm 16.3$  | 3 | $966 \pm 7.32$  | 12 |
| 9  | $1829 \pm 61.4$ | 3 | $1806 \pm 18.1$ | 3 | $1869 \pm 13.4$ | 3 | $1834 \pm 14.3$ | 9  |
| 10 | $260 \pm 10.0$  | 3 | $262 \pm 0.180$ | 3 | $258 \pm 4.69$  | 3 | $260 \pm 1.93$  | 9  |
|    | Silver (Ag)     |   |                 |   |                 |   |                 |    |
| 1  | $6.30 \pm 0.26$ | 4 | $6.66 \pm 0.09$ | 3 | $6.35 \pm 0.27$ | 4 | $6.30 \pm 0.06$ | 20 |
| 8  | $6.90 \pm 0.14$ | 3 | $6.79 \pm 0.16$ | 3 | $6.73 \pm 0.18$ | 3 | $6.81 \pm 0.04$ | 18 |
| 9  | $8.71 \pm 0.38$ | 3 | $8.51 \pm 0.28$ | 3 | $8.68 \pm 0.42$ | 3 | $8.66 \pm 0.08$ | 18 |
| 10 | $5.04 \pm 0.25$ | 3 | $5.21 \pm 0.09$ | 3 | $5.14 \pm 0.16$ | 3 | $5.04 \pm 0.05$ | 18 |

Some of the information needed to properly assess the value of Dr. Guinn's data and testimony relate to both compositional *homogeneity* and compositional *uniqueness* (individuality) of the Mannlicher–Carcano bullets. For example, how compositionally homogeneous (uniform) was each *batch* of bullets made by Mannlicher–Carcano (*homogeneity*), how compositionally homogeneous are the individual bullets derived from each batch (*homogeneity*), and how frequently do Mannlicher–Carcano bullets and bullets from other manufacturers "match" in composition (*uniqueness*).

Dr. Guinn testified about the homogeneity of Western Cartridge Company bullets to the House Select Committee on Assassinations [U.S. Cong. House (1979)], "One can only show what information we do have, and that is that you simply do not find a wide variation in composition within individual Western Cartridge Company Mannlicher–Carcano bullets, but you do find wide composition differences from bullet to bullet for this kind of bullet lead." This section further investigates this claim with regard to our study.

**4. Bullet homogeneity.** In our study we sampled the outer, middle and inner radial regions of each of four bullets. The four bullets were chosen based on preliminary measurements that best matched the assassination fragments. Table 3 gives the means and standard errors of antimony and silver measurements for the four bullets. Results in Table 3 are given at the radial portion of the bullet at which measurements were recorded. In addition, summaries for the entire bullet (which include measurements both labeled and unlabeled in terms of radial position) are reported in Table 3.

The differences in composition within a bullet are marginally detectable even with the small sample sizes used in our study. Our working hypothesis for this study is that bullets can be modeled as reasonably homogeneous in composition. However, in part because of variations in fabrication practices, and in light of the recent findings of Randich and Grant (2006), some bullets may not be homogeneous, disallowing homogeneity as a universal assumption. For example, when we compare silver from the middle and outer fragments of our bullet one from lot 6003, the p-value for comparing means using a pooled t-test (Fisher's LSD) is just less than 0.08 and when comparing arsenic for the middle and inner fragments for arsenic from the same bullet, the p-value is about 0.06. (Log-transformed data analysis produce similar p-values.) When a two-way multivariate analysis of variance on log silver and log arsenic with interaction term (factors "bullet" and "location") test is run using the three bullets from lot 6003, with replicates for both log silver and log arsenic, the p-value for bullet homogeneity is 0.04 (using Wilk's lambda; it is slightly lower using Hotelling–Lawley). These tests should be considered exploratory because of the small sample sizes and because other hypotheses were evaluated that did not show extraordinarily strong differences in homogeneity.

**5. Using the compositional bullet evidence.** To decide whether the compositional bullet lead evidence supports one shooter or more than one shooter, it is useful to use the ratio of probabilities $\frac{\Pr(T|E)}{\Pr(\bar{T}|E)}$, where $T$ denotes two bullets used in the assassination, $\bar{T}$ denotes its complement (in this case three or more bullets since at least two bullets were involved in the assassination), and $E$ stands for evidence. As in Carriquiry (2006), Bayes theorem can be used to obtain the equation

$$\frac{\Pr(T|E)}{\Pr(\bar{T}|E)} = \frac{\Pr(E|T)}{\Pr(E|\bar{T})} \frac{\Pr(T)}{\Pr(\bar{T})}.$$

The ratio on the left-hand side of the equation is a ratio of the probabilities for two bullets as the source for the five JFK fragments in question versus more than two bullets being the source for the five JFK fragments. The first ratio on the right-hand side is the ratio of probabilities for the evidence given two bullets versus having more than two bullets. This ratio allows us to decide whether, given the evidence, two bullets or more than two bullets is more likely.

Supporters of the lone-assassin interpretation, including Gerald Posner (1993), insist that Oswald missed the limousine occupants entirely with one of three shots, leaving only two of Oswald's bullets to wound President Kennedy and Governor Connally; therefore, a third shot striking both victims would require an additional shooter. Under this scenario, the second ratio on the right-hand side of the equation represents a ratio of the prior probabilities that we believe that there was a single shooter versus the fact that there are multiple shooters. This ratio is assessed independently of the compositional bullet measurements. In assessing the value of the evidence, characterizing the number of distinct compositional bullet groups in a box is important. To illustrate this point, we consider a situation similar to the outcome of our measurements from the ten bullet measurements from lot 6000. (These measurements are provided in the online supplemental materials.) To keep calculations as simple as possible and to emphasize that what is assumed about the number of matching bullets in a box matters, we assume no measurement error and no heterogeneity issues in the following calculation. We illustrate the calculation of the probability ratio by considering that the evidence of two bullets making up the five assassination fragments came from two or three bullets. We assume that there are two distinct groups of bullets from among ten measured bullets. This is similar to the groupings obtained for lot 6000 bullets using the $\pm$ four standard error criteria. We assume that there are six bullets in one group and four bullets in the other group. The hypergeometric distribution is an appropriate model distribution to use. If only two bullets were chosen, from our 10 bullets then there would be a 53.3% chance that two of them would come from two different groups. If three bullets were chosen, then there would be an 80% chance that they came from two groups. Thus, the critical ratio $\frac{\Pr(E|T)}{\Pr(E|\bar{T})} = \frac{0.53}{0.80}$. Since this ratio is less than 1, Dr. Guinn's testimony that the evidence supports two and only two bullets making up the five JFK fragments is fundamentally flawed, unless one places a very strong prior probability on $T$.

Dr. Guinn testified that essentially every bullet in a box of Mannlicher–Carcano bullets is unique and under that assumption, his conclusions are more logical. Our measurements on three different boxes of these bullets indicate a great many bullets have two element matches within a box. If more elements are measured and then compared for each bullet, there may be fewer within box matches. Also, if more elements are chemically analyzed, the JFK fragments may reasonably be placed into more than two groups. It is not known whether or not there was a second shooter. However, if there was a second shooter, the possibility exits that the bullets came from different boxes. The calculations above show that it is not possible from the compositional bullet lead analysis to conclude that there were only two bullets as the source of the five assassination fragments as Dr. Guinn testified. The answer could change dramatically depending upon assumptions. By reanalyzing the JFK bullet fragments to measure more elements, it may be possible to provide substantial evidence of more than two bullets as the source of the assassination fragments if there were, in fact, more than two bullets used. However, if bullets came from the same box, clear evidence of more than two bullets may not be present because many bullets in the same box typically have similar chemical compositions.

**6. Summary and conclusions.** We presented results from a study where ten bullets from each of three boxes of Mannlicher–Carcano bullets were analyzed for chemical composition. Compositional data from the ten bullets sampled from each box were compared to Dr. Guinn's testimony before the House Select Committee on Assassinations [U.S. Cong. House (1979)] regarding assassination bullet fragment compositions and also to the findings of the NRC in their report "*Weighing Bullet Lead Evidence*" (2004). We found that many bullets within a box of Mannlicher–Carcano bullets have similar composition. Further, we found that one of the thirty bullets analyzed in our study also compositionally matched one of the fragments from the assassination analyzed by Dr. Guinn [U.S. Cong. House (1979)]. If we allow for the bias associated with Dr. Guinn's measurements, it is possible that there would be even more matches among our bullets with the JFK fragments. We have shown that two-element chance matches to assassination fragments are not extraordinarily rare. Further, we have shown that if bullets come from the same box, they are even less rare. Given the significance and impact of the JFK assassination, it is scientifically desirable for the evidentiary fragments to be reanalyzed. The reanalysis should include at least the seven elements identified in the NRC report (2004), should establish the scientific basis for matching fragments originating from a single bullet, and should address the critically important issues of bullet and source heterogeneity.

## APPENDIX: EXPERIMENTAL DETAILS

The analyses performed in this study were designed to be comparable to the 1977 Neutron Activation Analysis (NAA) work of Guinn (1979). Variations in protocol were made to accommodate differences in reactors and counting instrumentation used by the different research teams. In the subsequent 30 years, significant advancements in detectors and electronics have occurred. Modern computing has facilitated rigorous peak extraction methods and analytical corrections. Still, the basic method used in both cases was very similar, and no development discounts the quality of the Guinn data.

The bullet lead matrix poses special concerns during NAA due to its heavy mass. The neutrons used as the analytical probe in this method find the lead sample to be fairly transparent, but the gamma-ray which is emitted from the subsequent decaying nuclide of the element of interest is likely to be attenuated. The issue is to ensure that any self-absorption [Knoll (1979)] of the gamma-rays is minimized and, if possible, corrected for. Since the likelihood of absorption is based on the path length of the emission through the sample as well as the absorption coefficient in lead, it is beneficial to use small samples.

Guinn used sample masses ranging from 1 to 50 mg, sizes dictated by the fragments available. The bullet recovered from Governor Connelly's stretcher (the "stretcher" bullet) was drilled and the resulting drillings/powder analyzed. In Guinn's study (1979), correcting for sample self-absorption because of variable geometry was difficult. In our study we used fragment chips prepared from the bullet lead material weighing 15 to 20 mg that had maximum linear dimensions between 0.1 mm and 1.5 mm, averaging 0.4 mm. Loss of gamma intensity by absorption should average about 2.47% to 3.34% for the range of gamma energies used at this average distance. Rather than risk inaccurate adjustments due to irregular specimen geometries, we included credible ranges for self-absorption in the calculation of the bias contribution to our uncertainty estimates.

Sample positioning in the neutron flux and counting positions is critical. The use of small samples to minimize self-absorption resulted in the need for additional measures to ensure that the sample was held at a precise location in the sample irradiation package. Sample portions were selected from the test bullets according to three different protocols. Three samples were prepared for each of 10 bullets from lot 6000 and a putative 6003 lot, for a total of 60 samples. We refer to this later set of bullets as coming from lot 6003P. Efforts were made to maximize the spread of locations sampled within a bullet. Additionally, ten bullets known to be from lot 6003 were sampled in a similar fashion, except three samples were removed from each of the three locations chosen for a total of nine samples per bullet, or 90 for the lot. Finally, several bullets (bullet numbers 1, 8, 9 and 10) from lot 6003 were sampled a second time. At each of the three sampling locations of these bullets, the three samples were taken from an interior slice with one sample representing the radial center of the bullet (inside), one representing the middle and one near the outside surface of the bullet (outside). Samples were prepared following Guinn's procedure (1979) and packaged into precleaned polyethylene vials.

Standardization—Commercial plasma solutions were used for comparator NAA standards.[2] Standards were prepared by depositing, and then drying, weighed portions of liquid solutions of known concentrations of the elements of interest into irradiation vials prepared exactly as for the bullet samples. Silver standards were prepared separately with approximately 2 micrograms of element deposited into each standard. Composite standards were prepared for arsenic, antimony and copper by depositing about 6, 19 and 12 micrograms respectively.

Quality material selection—Standard Reference Material Bullet Lead, SRM C2416 produced by the National Institute for Standards and Tech-

---

[2] Alfa Aesar, Specpure Plasma Standards, Ward Hill, MA.

nology (NIST)[3] is intended for use as a composition standard for optical emission spectroscopy (OES). The standard is not sectioned, but rather a newly exposed surface is presented to the instrumentation for analysis. Although NIST does not certify the homogeneity of the bulk standard nor provide a "minimum sample mass" as they do with standards intended for sub-sampling, and since no other reference material of similar matrix to our samples was available, the C2416 standard was the only feasible selection. We included three analyses of NIST SRM 1571, Orchard Leaves for arsenic,[4] as well as three analyses of a second independent set of plasma standards (referred to here as Trace Element Research Laboratory standards)[5] from a different manufacturer for silver, and six analyses for the other elements.

Irradiations and Counting—All irradiations were performed at the Texas A and M University's Nuclear Science Center 1 MW T.R.I.G.A. research reactor. The irradiation positions include a pneumatic tube and rotisseries, both with nominal neutron fluxes of $1 \times 10^{13}$ cm$^{-2}$ s$^{-1}$.

Silver was determined using the pneumatic facility by sequentially irradiating standards, unknowns and quality materials for 60 s ($T_i$), followed by a 30 s delay ($T_d$) and a 180 s count period ($T_c$). The $^{109}$Ag$(n,\gamma)^{110}$Ag reaction induced by thermal neutrons during the irradiation gave rise to the 657 keV gamma line ($E_\gamma$) from $^{110}$Ag which decays with a 24 s half-life ($t_{1/2}$). Measurements were taken using standard gamma-ray spectrometer protocols.

Arsenic, antimony and copper were each determined using longer-lived isotopes; therefore, a rotisserie irradiation of two hours was employed. Neutron capture reactions were responsible for production of indicator isotopes $^{76}$As ($t_{1/2} = 26.4$ h, $E_\gamma = 559$ keV), $^{122}$Sb($t_{1/2} = 2.70$ d, $E_\gamma = 564$ keV) and $^{64}$Cu($t_{1/2} = 12.7$ h, $E_\gamma = 511$ keV). Samples were counted in irradiation batches with each irradiation can being processed in consecutive counts on a single detector system.

Details of quality control material analyses are in supporting material but, in each case, the standard deviation of measurement distributions overlaps with the quoted uncertainty of NIST values. NIST specifies that the uncertainties for their certified standard values are not statistically derived measures of variability but rather are "based on judgment." Tabulated results of Trace Elements Research Lab standards are available in online supporting material.

---

[3]National Bureau of Standards, Certificate of Analysis, Standard Reference Material C2416, Bullet Lead, February 16, 1988, Gaithersburg, MD.

[4]National Bureau of Standards, Certificate of Analysis, Standard Reference Material 1571, Orchard Leaves, January 28, 1971, revised August 15, 1976, Gaithersburg, MD.

[5]Trace Element Research Laboratory Standards (TERL), Inorganic Ventures, Lakewood, NJ.

**Acknowledgments.** The authors are grateful to the Editors and an Associate Editor for many helpful comments that led to an improved paper.

# REFERENCES

CARRIQUIRY, A. (2006). Comment: "Further arguments against CABL as a forensic tool." *Chance* **19** 25–26.

GUINN, V. P. (1979). JFK assassination: Bullet analyses. *Anal. Chem.* **51** 484A–493A.

KNOLL, G. F. (1979). *Radiation Detection and Measurement*. Wiley, New York.

NATIONAL RESEARCH COUNCIL (2004). *Forensic Analysis*: *Weighing Bullet Lead Evidence*. The National Academies Press, Washington, DC.

PEELE, E. R., HAVEKOST, D. G., HALBERSTAM, R. C., KOONS, R. D., PETERS, C. A. and RILEY, J. P. (1991). Comparison of bullets using the elemental composition of the lead component. In *Proceedings of the International Symposium on the Forensic Aspects of Trace Evidence* 57–68. US Government Printing Office, Washington, DC.

POSNER, G. (1993). *Case Closed*. Bantam, New York.

RAHN, K. A. and STURDIVAN, L. M. (2004). Neutron activation and the JFK assassination. Part I. Data and interpretation. *J. Radioanalytical and Nuclear Chemistry* **262** 205–213.

RANDICH, E. and GRANT, P. M. (2006). Proper assessment of the JFK assassination bullet lead evidence from metallurgical and statistical perspectives. *J. Forensic Sci.* **51** 717–728.

SPIEGELMAN, C. H. and KAFADAR, K. (2006). Data integrity and the scientific method: The case of bullet lead data as forensic evidence. *Chance* **19** 17–24. MR2247019

STURDIVAN, L. M. and RAHN, K. A. (2004). Neutron activation and the JFK assassination. Part II. Extended benefits. *J. Radioanalytical and Nuclear Chemistry* **262** 215–222.

TOBIN, W. A. and THOMPSON, W. C. (2006). Evaluating and challenging forensic identification evidence. *The Champion NACDL* 12–21.

U.S. CONGRESS HOUSE, SELECT COMMITTEE ON ASSASSINATIONS. (1979). *Report of the Select Commitee on Assassinations of the U.S. House of Representatives*. 95th Cong., 2nd Sess., House Report No. 95-1828, Part 1-b, Washington, GPO.

C. SPIEGELMAN  
S. J. SHEATHER  
DEPARTMENT OF STATISTICS  
TEXAS A&M UNIVERSITY  
3143 TAMU  
COLLEGE STATION, TEXAS 77843-3143  
USA  
E-MAIL: cliff@stat.tamu.edu

W. A. TOBIN  
FORENSIC ENGINEERING INTERNATIONAL  
2708 LITTLE GUNSTOCK RD.  
LAKE ANNA, VIRGINIA 23024-8882  
USA

W. D. JAMES  
CENTER FOR CHEMICAL CHARACTERIZATION  
AND ANALYSIS  
TEXAS A&M UNIVERSITY  
3144 TAMU  
COLLEGE STATION, TEXAS 77843-3144  
USA

S. WEXLER  
HUMANITIES AND ADVANCED  
PLACEMENT GOVERNMENT  
HIGHSTOWN HIGH SCHOOL  
25 LESHIN LANE  
HIGHTSTOWN, NEW JERSEY 08520  
USA

D. M. ROUNDHILL  
CHEM CONSULTING  
13325 BLACK CANYON DRIVE  
AUSTIN, TEXAS 78729  
USA